\documentclass[reprint,aps,pra, 10pt,
superscriptaddress,
amsmath,amssymb,
floatfix
]{revtex4-1} % For final two-column format

\usepackage{amsmath,amssymb,bm}
\usepackage{array}
\usepackage{booktabs}
\usepackage{esvect}
\usepackage{graphicx}
\usepackage{mathtools}
\usepackage{multirow}
\usepackage{comment}
\usepackage{hyperref}

\usepackage[svgnames]{xcolor}
\begin{document}

\title{Searching for New Interactions at Sub-micron Scale Using the M\"ossbauer Effect}

\author{Giorgio Gratta}
\affiliation{Department of Physics, Stanford University, Stanford, California 94305, USA}

\author{David E. Kaplan}
\affiliation{Department of Physics \& Astronomy, The Johns Hopkins University, Baltimore, MD  21218, USA}

\author{Surjeet Rajendran}
\affiliation{Department of Physics \& Astronomy, The Johns Hopkins University, Baltimore, MD  21218, USA}

\date{\today}
\begin{abstract}

A new technique to search for new scalar and tensor interactions at the sub-micrometer scale is presented.   The technique relies on small shifts of nuclear $\gamma$ lines produced by the coupling between  matter  and the nuclei in the source or absorber of a M\"ossbauer spectrometer.  Remarkably, such  energy shifts are rather insensitive to electromagnetic interactions that represent the largest background in searches for new forces using atomic matter.  This is because nuclei are intrinsically shielded by the electron clouds. Additionally, electromagnetic interactions cause energy shifts by coupling to nuclear moments that are suppressed by the size of the nuclei, while new scalar interactions can directly affect these shifts.  Finally, averaging over unpolarized nuclei, further reduces electromagnetic interactions.   We discuss several possible configurations, using the traditional M\"ossbauer effect as well as nuclear resonant absorption driven by synchrotron radiation.  For this purpose, we examine the viability of well known M\"ossbauer nuclides along with more exotic ones that result in substantially narrower resonances.   We find that the technique introduced here could substantially improve the sensitivity to a variety of new interactions and could also be used, in conjunction with mechanical force measurements, to corroborate a discovery or explore the new physics that may be behind a discovery.
\end{abstract}

\maketitle

\section{Introduction}
\label{sec:intro}

Light, weakly coupled particles emerge in many theories of physics beyond the standard model. Examples of such particles include scalars such as moduli~\cite{Adelberger:2003zx} and relaxions~\cite{Graham:2015cka, Graham:2017hfr,Graham:2019bfu} that are tied to solutions of the hierarchy problem as well as mediators between the standard model and the dark sector. These particles can be experimentally probed by searching for the new forces they mediate between standard model particles. Current limits~\cite{Adelberger:2003zx, IUPUI} on such forces are a strong function of the range ({\it i.e.} mass) of the new particle. Roughly, for distances greater than tens of microns, the strength of the new force is constrained to be weaker than gravity. But, at distances below the micron scale, forces that are many orders of magnitude larger than gravity are still allowed. 

There are two fundamental reasons for this sharp dependence of the sensitivity on the range. First, as the range decreases, only a progressively thinner sliver of material is at the correct distance to probe the interaction, so that the force becomes weaker. Second, and more importantly, in the case of experiments based upon the interaction between atomic matter~\cite{experiments}, electromagnetic effects such as Casimir forces~\cite{Casimir} and gradient interactions, such as produced by patch potentials, increase rapidly at short distances, resulting in overwhelming backgrounds.  While conductive shields can mitigate both effects, practical considerations related to the small distances involved and the finite conductivity of materials limit their effectiveness in the very short range regime.  Indeed, while to-date no experiment using interactions between atoms has reported the discovery of a new force at the sub-mm scale, it is reasonable to ask whether these techniques would offer sufficient redundancy and cross checks to support a positive claim of extraordinary importance.   The use of neutrons, with a charge radius that is much smaller than that of atomic matter, has also been pursued in scattering experiments~\cite{neutrons}. Although in this case the systematics are very different, the sensitivity is substantially lower due to the difficulty in obtaining suitable neutron sources.

%This background currently limits the use of atom and optical interferometer techniques in probing sub-micron forces. One approach that has been tried to circumvent this problem is to use a ``casimir shield'' where a common material layer is placed between different material sources and the probe with the hope that the shield will make the casimir effects sufficiently uniform and thus cancel out in a differential measurement. While this is an interesting possibility and could be useful in setting limits on such a force, it is unclear if this approach could lead to a clear discovery. 

It is thus interesting to investigate sensing platforms that might naturally suppress the large electromagnetic backgrounds while retaining sensitivity. In this paper, we point out that there is an enticing possibility to probe the direct coupling of nuclei to scalar and tensor interactions.   This can be achieved by studying, by means of M\"ossbauer spectroscopy~\cite{Mossbauer}, the very small energy shifts expected when nuclei are exposed to new scalar and tensor interactions.  The magnitude of such a shift only depends upon the strength of the interaction and is unsuppressed by nuclear moments.  At the same time, electromagnetic effects on nuclear energy levels are significantly suppressed, being shielded by the electron clouds. Moreover, electromagnetic effects can shift nuclear energy levels only through multipole effects, and these are suppressed by the size of the nucleus. Further, these multipole moments are set by the spin of the nucleus. In a sample where the spins are not aligned, these effects will average down, unlike the effects of the signal from a new scalar  interaction. 
%The M\"ossbauer effect thus offers an interesting path towards enhanced sensitivity while simultaneously reducing electromagnetic backgrounds in searching for new scalar  particles. 
We note that this method will not be useful in searching for new vector forces, since those are analogous to electromagnetism, and their effect on nuclear energy levels will be suppressed accordingly.   One may take the point of view that such a limitation can actually be exploited to measure the spin of the new mediator by comparing a search with this technique with one using atomic matter.

In the following, we investigate different M\"ossbauer setups and sources to search for such scalar and tensor forces. We begin in section~\ref{sec:concept} with a conceptual overview of the setup, estimating the likely systematics. We then discuss three possible experimental realizations in Section~\ref{sec:experiment}.  In section~\ref{sec:reach} we estimate the effects of the new interactions on nuclear energy levels and compute the potential reach of the experimental approach. Finally, in section~\ref{sec:conclusions}, we conclude.

\section{Concept}
\label{sec:concept}

\begin{figure}
	\includegraphics[width=0.95 \linewidth]{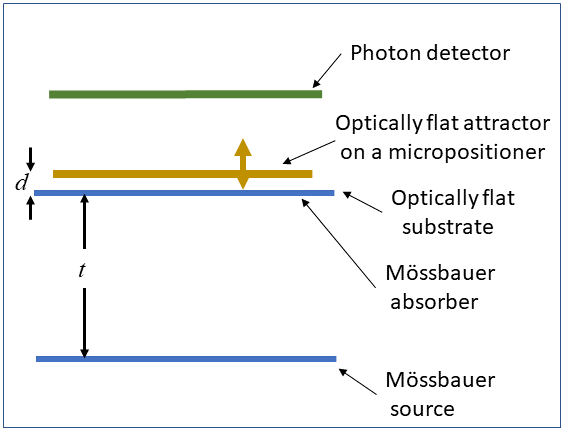}
	\caption{Conceptual sketch of the type of experiment proposed.  The distance $d$ between attractor and absorber can be adjusted to probe the effect of a new interaction.  The attractor may not even be a self-supporting foil but, rather, a layer of a solid grown onto the absorber.  The distance $t$ between source and absorber and the exact geometry of the two M\"ossbauer foils and of the photon detector are inessential and can be optimized differently for different experimental realizations. }
	\label{fig:sketch}
\end{figure}

In the scheme proposed here, generically illustrated in Figure~\ref{fig:sketch}, photons are emitted by a M\"ossbauer source. The resonant re-absorption of these photons by an absorber is tested when the source or the absorber are perturbed by an ``attractor'' that generates the new interaction. A change in the re-absorption cross-section as a function of the distance $d$ between the ``attractor'' layer and the source or the absorber would reveal the existence of the new interaction. 

In the rest of the paper, we will use the terms ``source'' and ``absorber'', even if, in some cases, the two maybe identical. Further, the device generating the new interaction will be called ``attractor'' irrespective of the sign of the interaction. While, at least in principle, the attractor can be setup to perturb either the source or the absorber, it will arbitrarily assumed that the absorber is perturbed.  The choice between the two possibilities will depend on the technical details of a design.  Typically the new interaction is applied across a planar gap, although other geometries are possible.   Likewise, in Figure~\ref{fig:sketch}, the effect is notionally illustrated by measuring the absorption of photons with a detector that is co-planar with the absorber, but other schemes whereby the incoherent re-emission from the absorber is used are also possible.  The attractor can be though of as a self-supporting slab, as illustrated in Figure~\ref{fig:sketch}, positioned at variable distances from the absorber by means of piezoelectric actuators.  In this fashion one can plausibly adjust the distance down to a fraction of a micron, for properly planarized surfaces in vacuum.  However, in a different scheme, a solid layer can be condensed on the surface of the absorber, with the thickness of the layer setting the distance scale probed.  Xenon may be an ideal choice for this application, owing to the high freezing temperature and large atomic mass.  The use of separated isotopes would also allow to test the influence on the measurement of the number of neutrons and the nuclear spin.  

The relationship between the new force and the energy shift, $\Delta E$, is discussed in Section~\ref{sec:reach}.  The resonant absorption cross-section of a photon of energy $E_{\gamma}$ is given by
\begin{equation}
\sigma_0(E_{\gamma}, E) = \frac{2 \pi}{E^2} \frac{1}{1 + \beta} \frac{2 I' + 1}{2 I + 1} \frac{\left(\Gamma/2\right)^2}{\left(E - E_\gamma\right)^2 + \left(\Gamma/2\right)^2}
\label{eq:sigma0}
\end{equation}
\noindent where $\Gamma$ and $E$ are, respectively, the natural line width and  energy of the resonance, $\beta$ the internal conversion coefficient and $I$ ($I'$) the spin of the ground (excited) state.  The observed width may be larger than the natural one, because of non-homogeneous condensed matter conditions around the nuclei of interest. Such a spread is often modeled as a Gaussian function, to be convolved with the expression \eqref{eq:sigma0}.   It is clear that, for the present purpose, the system should approximate the natural line width as closely as possible (and the nuclides with the narrowest $\Gamma$ provide the best probes). Apart from line broadening, in some cases condensed matter effects may result in line shifts that may be different in sources and absorbers.  These play no role in the measurements proposed here, since they can be calibrated-out with the attractor in a retracted position.

Sensitivities to $\Delta E$ in the range $10^{-15}$~eV to $10^{-17}$eV are found in Section~\ref{sec:reach} to be achievable from the line widths and counting statistics in systems using  known technology. Here we examine the limits imposed on these measurements by the backgrounds produced by electromagnetic coupling to the attractor.  These would arise as shifts of the transition energies, $\delta E_{bkgd}$, that cannot be easily distinguishable from the shift, $\Delta E$, due to the signal.  $\delta E_{bkgd}$ for a single nucleus, can be estimated for Casimir forces, patch potentials and magnetic impurities.  The electric field due to the Casimir effect between two plates that are separated by a distance $d$ is $\delta E_{bkgd}^{\rm Casimir} \approx \sqrt{\frac{4 \pi}{240}}\frac{\pi}{d^2}$, so that the resulting energy shift to a nucleus through the electric quadrupole moment is $\approx  \sqrt{\frac{4 \pi}{240}}\frac{\pi \alpha_{\text{EM}} r_N^2}{d^3} $ where $r_N$ is the size of the nucleus and $\alpha_{\text{EM}}$ is the fine-structure constant. Taking, conservatively, $r_N $ to be the radius of the tantalum nucleus and $d \approx 10^{-8}$ m, an energy shift $\delta E_{bkgd}^{\rm Casimir} \approx 10^{-13}$~eV is obtained. Electric fields from electrostatic patch potentials roughly scale as $V_{0}/d$ where $V_0 \approx 100$~mV for distances $d \approx 10$~nm~\cite{Behunin:2011gj}, resulting in a shift $\delta E_{bkgd}^{\rm patch}\approx 10^{-15}$~eV. Finally, ferromagnetic domains in the attractor should contain fewer than $\approx 100$ polarized spins in the 10~nm scale when the attractor is at $d \approx 10^{-8}$~m, inducing line shifts $\delta E_{bkgd}^{\rm Mag} < 10^{-10}$~eV.    Since, at first order, these backgrounds couple to magnetic dipole and electric quadrupole moments of nuclei, for unpolarized samples the overall shift will be averaged down as $\delta E_{bkgd} / \sqrt{N_{\gamma}}$, where $N_{\gamma}$ is the number of nuclei participating in the resonant absorption experiment (i.e. the number of events in the experiment). Typical experiments discussed in the following use $N_{\gamma} \gtrapprox 10^{13}$, so that, even the largest background shift discussed here, $\delta E_{bkgd}^{\rm Mag}$, is expected to average down to a sufficient level not to limit the sensitivity.  The nucleus-to-nucleus difference in $\delta E_{bkgd}$ will result in a small line broadening, however such broadenings should be measurable independently from the energy shifts. 

Second order electromagnetic shifts can arise through the chemical/isomeric shift in the nucleus. Unlike the first order effects discussed above, these do not average down with the number of events and arise due to the overlap between the electron clouds and the finite size of the nucleus. The intrinsic value of the resulting shift is dominated by the inner electrons that are closer to the nucleus. This shift becomes a background to the measurement only if its magnitude changes with the distance between the attractor and the absorber, as it can be the case when the electric field from the attractor polarizes the electron clouds. It can be verified that this effect is dominated by the outermost electrons. The contribution of the outermost electron clouds to chemical/isomeric shift is $\approx 5 \times 10^{-9}$~eV for iron-like elements~\cite{ChemicalShift, Akai}. An electric field $E_0$ will change the overlap of the outermost s-electron by $\approx  \alpha_{\text{EM}} \left(E_0 a_B/\omega_e\right)^2$ where $a_B$ is the Bohr-radius and $\omega_e$ the binding energy of the electron. Taking typical values $a_B \approx 0.05 $~nm and $\omega_e \approx 10$~eV, it is estimated that this energy shift is dominated by Casimir forces and is $\approx 3 \times 10^{-15} \left(\frac{10 \text{ nm}}{d}\right)^4$~eV. It is likely that smaller shifts than this can be obtained with a judicious choice of the chemistry of the absorber. For example, if the absorber is made of an ionic compound of iron where iron is doubly oxidised, its outermost s-electrons would have a reduced overlap with the nucleus. Since these electrons are the ones that are most easily polarized, the shift is likely to decrease. Given the uncertainties, we will conservatively take this line shift to be $10^{-14} \left(\frac{10 \text{ nm}}{d}\right)^4$~eV. While this number was calculated for iron, due to the dependence of the effect on chemical properties of the absorber,  we will use it as a limit to the sensitivity of other nuclei as well. At 10~nm, this shift is larger than the statistical sensitivity that could potentially be reached by the experiment, but since it drops rapidly with distance, it quickly becomes sub-dominant. Note that lattice imperfections can give rise to first order effects wherein defects cause an intrinsic polarization of the electron cloud which can then couple to these electric fields at first order. However, these do average down and are sub-dominant to the effects discussed above. 

Two types of temperature effects need to be considered in analysing systematics for the proposed measurements~\cite{SOD_LambMossbauer}: the second order Doppler shift and the Lamb-M\"ossbauer effect. The second order Doppler shift can be understood as the dilation of proper time for nuclei undergoing thermal vibrations in the lattice, resulting in a shift, $\delta E_{temp}$, dependent on the temperature. Unlike the linear counterpart, this effect does not average out in the lattice. The temperature dependence~\cite{Dunham, Nasu} is $\approx 10^{-11} \left(T/\text{300 K}\right)^3$~eV/K, resulting in shifts that can be caused by temperature drifts. There are two possible strategies to mitigate these effects. The first is to run the experiment at low temperatures ({\it e.g.}$\approx 4$ K), where the value of the shift is smaller than the line shifts of interest. The second consists in monitoring the temperatures of the source and the absorber and correct for the effect. A combination of these approaches could also be pursued. For example, if the system is cooled to $\approx 30$~K, the average temperature of the system needs to be known to within 100~mK over the course of the measurement to keep systematic shifts below $10^{-15}$~eV. An additional consideration that arises from the second order Doppler shift is the possibility that the Debye temperature of the lattice may be changed by electromagnetic backgrounds when the attractor is brought near the absorber. For the electric field to change the Debye temperature, it has to cause relative motion between the atoms in the lattice, analogous to a tidal effect. Further, since the atoms are neutral, the leading order force on them is through a dipole moment. This effect should also be suppressed by the mass of the nucleus and the intrinsic stiffness {\it i.e.} the Debye frequency of the system. Thus, the fractional change to the line width is at most  $\delta E /E \approx \left(\mathcal{E_0} \left(\frac{d_{a}}{d}\right)^2 \frac{1}{M \theta_{D}} \right)^2$ where $\mathcal{E_0}$ is the electric field over a distance $d$, $d_a$ the electric dipole moment of the atom, $M$ the mass of the nucleus and $\theta_D$ the Debye temperature of the lattice. In deriving this estimate, we have made use of the fact that terms that are linear in the intrinsic ``velocity'' of the nucleus in the lattice do not cause line shifts. It can be verified that this effect is significantly smaller than the second order chemical shift discussed above. 

The second effect of the temperature is through the Lamb-M\"ossbauer factor which changes the fraction of recoilless emission from the source. While this is not a line shift, it will change the number of resonant events, simulating a shift of the emission or absorption lines.  At temperatures $T \ll \theta_D$,  this fraction is $\text{e}^{-3 \frac{E}{\theta_D}\left(1 + \frac{2 \pi^2 T^2}{3 \theta_D^2} \right)}$~\cite{LambMossbauer}. For M\"ossbauer emissions $E\approx 10$~keV,  $\theta_D \approx 400$~K and $T \approx 30$~K, the temperature coefficient of the resonant rate is $\approx 10^{-4}$~K$^{-1}$.  For this effect not to exceed the statistical fluctuations over the $10^{13}$~events/month considered in Sec.~\ref{sec:reach}, is sufficient to limit or measure temperature fluctuations at the 3~mK level.  This systematic can also be mitigated by continuously normalizing the resonant rate, e.g. using two absorbers, one of which is not perturbed by the attractor.

\section{Experimental Realization} 
\label{sec:experiment} 

We investigate the potential of traditional sources with modest lifetimes and natural line widths $10^{-11} \lesssim \Gamma \lesssim 10^{-9}$~eV.  Those are listed above the line in Table~\ref{tab:properties}.  For these nuclides, the traditional M\"ossbauer effect has been experimentally demonstrated with line widths comparable to the natural ones.  As discussed in Section~\ref{sec:reach}, we find that traditional M\"ossbauer experiments based on these nuclides can be used as competitive probes for new forces with range as short as 100~nm and, possibly, shorter.   

Nuclides in the second part of Table~\ref{tab:properties} could plausibly further improve the sensitivity of the technique, owing to the exceedingly narrow natural line widths. For this reason, these nuclides may be also competitive with force measurements for distances larger than $\gtrapprox 100$~nm. However, substantial work is required to take advantage of these very sharp resonances, possibly in conjunction with the use of synchrotron radiation for their excitation.

\begin{table}[t]
\renewcommand{\arraystretch}{1.2}
    \begin{tabular}{ccccc}
Nuclide &   $E$ (eV) & $T_{1/2}$ & $\Gamma$ (eV) &  $\Gamma /E$    \\
\\[-8pt]
    \hline
    \vspace{-0.2cm} \\
$^{57}_{26}$Fe  &  14,413   & 98.3 ns     & $4.7\times 10^{-9}$  & $6.4\times 10^{-13}$  \\
$^{73}_{32}$Ge  &  13,328   & 2.92 $\mu$s & $1.6\times 10^{-10}$  & $1.2\times 10^{-14}$  \\
$^{181}_{73}$Ta &  6,237    & 6.05 $\mu$s & $7.5\times 10^{-11}$ & $1.2\times 10^{-14}$  \\
$^{67}_{30}$Zn  &  93,300   & 9.07 $\mu$s & $5.0\times 10^{-11}$  & $5.4\times 10^{-16}$  \\
\\[-8pt]
\hline
\\[-8pt]
$^{45}_{21}$Sc  &  12,400  & 318 ms      & $1.4\times 10^{-15}$  & $1.13\times 10^{-19}$  \\
$^{107}_{47}$Ag &  93,125  & 44.3 s      & $1.03\times 10^{-17}$ & $1.1\times 10^{-22}$  \\
$^{103}_{45}$Rh &  39,753  & 56.1 min    & $1.36\times 10^{-19}$ & $3.4\times 10^{-24}$  \\
$^{189}_{76}$Os &  30,814  & 5.8 hr      & $2.2\times 10^{-20}$  & $7.0\times 10^{-25}$ \\
\\[-8pt]
    \hline
    \end{tabular}
\caption{Properties of some nuclides of interest~\cite{BNL} ordered by the half life of the M\"ossbauer transition $T_{1/2}$. $E$ and $\Gamma$ are the energy and the natural line width of such transition, the latter calculated from the half-life. The four nuclides above the line have relatively short half lives and are mostly discussed in the context of traditional M\"ossbauer setups, while nuclides below the line are though of as more aggressive options, requiring substantial R\&D and the use of excitation by synchrotron radiation. }
    \label{tab:properties}
\end{table}

\begin{table*}[th!!!!!!!!!!!]
\renewcommand{\arraystretch}{1.2}
    \begin{tabular}{ccccccccc|ccc}
    \multicolumn{9}{c}{M\"ossbauer Decay}  & \multicolumn{3}{|c}{Parent Properties}           \\
    \hline
N$_{\rm M}$ & $E$ &  $\Gamma$  &  $\Gamma_{\rm EXP}$ & $\Gamma_{\rm EXP}/\Gamma$ & Ref. & $\beta$ & $\ell$  & $\eta$ & N$_{\rm P}$ & Decay & Half   \\
            &     (eV)           &           (eV)               &         (eV)                  &      &      &    & (nm)  &   & & mode  & life (d)  \\
    \hline
    \vspace{-0.45cm} \\
%$^{45}_{21}$Sc  & $1.4\times 10^{-18}$ &     &  &                                              & $^{45}_{20}$Ca & $\beta^-$ & 162.6 & $1.9\times 10^{-3}$ \\
$^{57}_{26}$Fe  & 14,413 & $4.7\times 10^{-9}$  & $\approx 5\times 10^{-9}$ & $\approx 1$ & \cite{cyclotron}& 8.56 & 48 & 0.89    & $^{57}_{27}$Co & EC & 272   \\
$^{73}_{32}$Ge & 13,328 & $1.6\times 10^{-10}$ & $1.6\times 10^{-10}$  & $\approx 1$ & \cite{Pfeiffer_77} & $1.12\times 10^3$ & $3.2\times 10^4$ & 1  & $^{73}_{33}$As & EC & 80.3  \\
$^{181}_{73}$Ta & 6,237 & $7.5\times 10^{-11}$ & $5.5\times 10^{-10}$      & 7.5    & \cite{Dornow} & 70.5 & 180 & 1      & $^{181}_{74}$W & EC & 121.2  \\
$^{67}_{30}$Zn & 93,300 & $5.0\times 10^{-11}$ & $7.5\times 10^{-11}$     & 1.5    & \cite{KatilaRiski} & 0.87 & $3.3\times 10^3$  & 1 & $^{67}_{31}$Ga & EC & 3.2     \\
    \hline
    \end{tabular}
\caption{For the first four nuclides $\rm N_M$ in Table~\ref{tab:properties}, left columns: decay energy ($E$), natural line width $\Gamma$ and line width $\Gamma_{\rm EXP}$ derived from data reported in the references listed.     In the case of $^{57}$Fe, $\Gamma_{\rm EXP}$ is that of a commercial supplier, indicated in the reference. $\beta$ is the internal conversion coefficient, as already mentioned, $\ell$ is the resonant mean free path, in a sample of pure $\rm N_M$, calculated from Eq.~\ref{eq:sigma0}, and $\eta$ is the fraction of decays of the parent nuclide $\rm N_P$ landing in the M\"ossbauer state.  Right columns: properties of the commonly used parent nuclides.
}
    \label{tab:Traditional}
\end{table*}

\subsection{Traditional M\"ossbauer Technique} 
\label{sec:traditional}

In the traditional M\"ossbauer technique, the appropriate nuclide N$_{\rm M}$ is produced directly in the isomeric state N$_{\rm M}^*$ from a different progenitor nuclide N$_{\rm P}$, using a beta or EC decay with a convenient half-life. Hence, the source is intrinsically non-homogeneous.  The absorber can be entirely made of the nuclide N$_{\rm M}$ in its ground state so that, typically, there is an energy shift between emission and absorption lines.  More parameters for the four nuclides at the top of Table~\ref{tab:properties} are shown in Table~\ref{tab:Traditional}, where the natural line widths are compared with experimentally obtained values (in the $^{57}$Fe case by commercial sources).     In our application, the attractor has to be brought in close proximity to the source or the absorber.  Since source preparation is complex, here we concentrate on the absorber that needs to be thick enough to provide full resonant absorption, to achieve the best constrast, while, at the same time, no thicker than the range at which the new interaction is tested.  The resonant mean-free path, $\ell$, can be derived from Eq.~\ref{eq:sigma0}. An efficient absorber for $^{57}$Fe can be made by coating some inert substrate with a layer sufficiently thin to reach attractor distances below 100~nm.  From the values of $\ell$ in Table~\ref{tab:Traditional}, it appears that similar properties can be achieved with $^{181}$Ta using a few-layer array of absorbers and attractors. $^{73}$Ge and $^{67}$Zn are not suitable for the study of short distance interactions, unless an arrangement can be found to apply the attractor to the source.

In order to maximize the statistical power of the measurement, it is also important to use a system in which the product $(\frac{1}{1+\beta}) \eta$ is maximized. Here $\eta$ is the fraction of decays of the progenitor feeding the M\"ossbauer state.  Since low energy transitions have larger internal conversion coefficients, the desire for a large $\frac{1}{1+\beta}$ is at odds with the other requirement that the M\"ossbauer transition should have low energy, so that a substantial fraction of decays is recoilless at room temperature.  The recoilless fraction is not shown in Table~\ref{tab:Traditional} because it depends on the matrix of the source and the absorber, but for $^{57}$Fe, $^{73}$Ge and $^{181}$Ta in metallic matrices can be assumed to be around 70\% at 300~K.   The M\"ossbauer transition in $^{67}$Zn has a much larger energy and hence a substantial recoilless fraction can only be achieved at low temperature, as was done in~\cite{KatilaRiski}.  

In an ideal experiment, similar statistics would be acquired without attractor, with source and absorber in perfect resonance, and with the attractor potentially shifting the line.  Because of the shift between source and absorber due to the different matrices, some mechanism to scan the lines is required.   This is usually achieved using Doppler drive systems where constant velocity is often achievable for the majority of the stroke.  An alternative technique may consist in producing a shift with an external magnetic field, which can be properly tuned and operated statically during data taking.  The shift produced in this way is $\approx 10^{-7}$~eV/T.  However, while this solution may be ideal when the excitation is provided by synchrotron radiation and only sub-line width shifts are required, when different matrices in the source and absorber require larger shifts, the efficiency may be reduced because magnetic splitting will produce several lines, only one of which can be used in the measurement.  

From the discussion above, it appears that the most common source, $^{57}$Fe, is also best suited for the technique described here, using a traditional M\"ossbauer setup.  In section~\ref{sec:reach} the sensitivity is computed for the cases of $^{57}$Fe using an activity of 100~mCi, and a $\pm 10^{\circ}$ collimation, resulting in a 0.03~sr solid angle and a 6\% line broadening, as would be the case for the usual Doppler tuning.

\subsection{Coherent Synchrotron Light Excitation} 
\label{sec:coherent}

Since the advent of synchrotron light sources, the possibility has arisen of directly exciting the isomeric state from the ground state by the absorption of a photon~\cite{synch_first, synch_review}.  For the purpose of interest here, this has the advantage that source and target can be made out of the same material and resonant absorption is achieved automatically, with a minimum amount of line shift required to scan the resonance and probe the signal from a new interaction.  Hence, the use of a magnetic field to shift the line, as mentioned above, is expected to be the technique of choice.  More importantly, the use of excitation by synchrotron radiation opens the possibility of using nuclides for which there is no obvious production scheme with the traditional method.   Because of the homogeneous nature of the source, it is also possible that ultra-sharp lines, such as those resulting from the nuclides in the second part of Table~\ref{tab:properties}, can be eventually approached.

Other peculiar properties of synchrotron light excitation may not be relevant here.   While in all cases M\"ossbauer spectroscopy relies on the coherent recoil of an entire crystal, the radiation emitted after synchrotron light excitation is also coherent in the sense of superradiance~\cite{superradiance}, whereby the emission can be considered as a wave common to the entire crystal.  In this picture, the exponential decay of the isomeric state should be thought of as the decrease in the amplitude of the wave and sufficiently large energy shifts between source and absorber manifest themselves in ``quantum beats'' modulating the exponential decay.  In practice, energy shifts at the threshold of detection, generally contemplated in the measurements discussed here, do not produce appreciable modulation and their detection has to be based on the more mundane overall rate decrease.   Furthermore, the drastic background reduction afforded by the fast synchrotron pulses, allowing for the observation of the resonance as nuclei de-excite after the end of the pulse, is only applicable to cases where the lifetimes are shorter than the repetition rate of the synchrotron light.  

The main limitation of the the excitation by synchrotron radiation derives from the achievable statistics.  This is because of the substantial mismatch between the spectral density of the synchrotron radiation and the exceedingly sharp absorption line of the isomeric transitions.   Dedicated monochromators exist at least at the APS (Argonne National Lab, USA)~\cite{APS} and SPring-8 (Japan)~\cite{SPring-8} with $\approx 1$~meV linewidth and peak intensities of $5\times 10^{13}~\gamma$/s at 14.4~keV (BL09XU beamline of SPring-8).   With this state of the art equipment, $5\times 10^{-6}$ ($1.4\times 10^{-12}$) of the available photons are useful to excite the $^{57}$Fe ($^{45}$Sc) transition, resulting in an integrated rate of $10^{14}$~month$^{-1}$ ($6\times 10^5$~month$^{-1}$), after accounting for the gamma conversion factor $\beta$. The first figure already exceeds the rate of $10^{13}$~month$^{-1}$ expected for the 100~mCi traditional source collimated to $\pm 10^{\circ}$ discussed in the previous section. While an experiment at a synchrotron radiation facility is more involved than a traditional M\"ossbauer one, shorter run times, consistent with typical beam line usage appear realistic and the possibility of magnetic tuning may represent an appealing option.  While an alternative approach to further increasing the resonant event rate is offered in the following section, it is also possible that the advent of proper X-ray lasers~\cite{Xray_laser} will result in sources with narrower band and higher spectral density.

\subsection{Three-State Synchroton Light Excitation} 
\label{sec:absorption}

\begin{figure}
	\includegraphics[width=0.75 \linewidth]{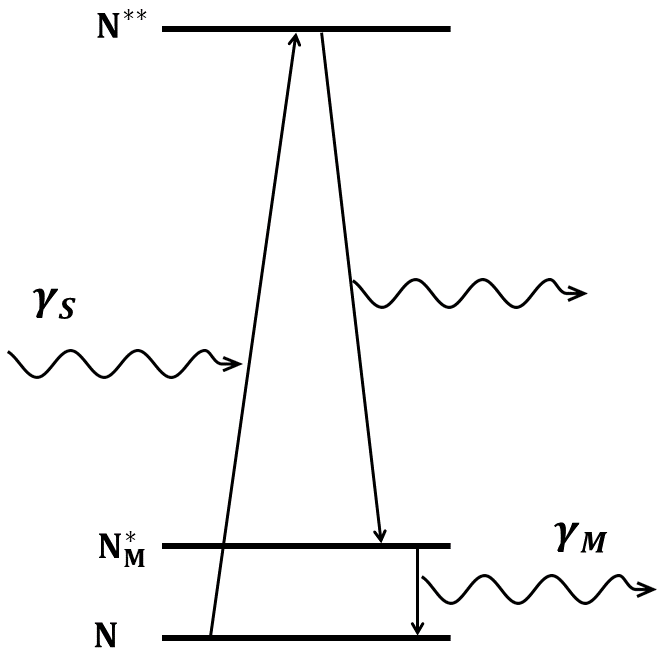}
	\caption{Three level scheme discussed in the text.  The ground state $\mathrm N_M$ is excited to $\mathrm N^{**}_M$ by a (relatively) broad-band synchrotron light photon, $\gamma_S$.  $\mathrm N^{**}_M$ then de-excites to the very narrow M\"ossbauer state $\mathrm N^{*}_M$, which decays back to the ground state with the emission of the photon $\gamma_{\mathrm M}$. }
	\label{fig:3level}
\end{figure}

The poor coupling between the energy spectrum of the synchrotron and the isomeric transition may be mitigated using a three-level system, as illustrated in Figure~\ref{fig:3level}.   The higher energy state N$^{**}$ is excited by synchrotron radiation with good efficiency, owing to its substantial width.  N$^{**}$ then spontaneously decays into the M\"ossbauer state of interest, N$^*_M$.  While this process does not involve the coherence mentioned above and, hence, it is not expected to produce quantum beats, as discussed this is not important for the current application. 

The mechanism requires a state $\rm N^{**}$ with sizeable amplitudes connecting it to both $\rm N^{*}_M$ and the ground state N. This appears to be the case at least for $^{189}$Os, with a $\rm N^{**}$ state of $J^P = 7/2^-$, energy 216.67~keV and half life 0.4~ns.  The excitation energy is challenging but not out of reach for synchrotron radiation sources.  Importantly, the strength of the transition $\rm N^{**}\rightarrow N^*_M$, $k^* = 100$, is comparable to that of the transition $\rm N^{**}\rightarrow N$, $k_0 = 34.3$, meaning that the state $\rm N^{**}$ is accessible with similar probability from the ground and the M\"ossbauer states.  The gain in rate with respect to direct synchrotron radiation excitation would be given by 
\begin{equation}
    A \approx \frac{\Gamma^{**}}{\Gamma} \frac{\mathcal I^{**}}{\mathcal I^*_M} \frac{k^* k_0}{(k^* + k_0)^2}
\end{equation}

\noindent where $\Gamma^{**}$ is the line width of the $\rm N^{**}$ state and $\mathcal I^*_{\rm M}$ ($\mathcal I^{**}$) the spectral density of synchrotron radiation source for the transition energy from the ground state to $\rm N^*_M$ ($\rm N^{**}$).  Using parameters for the BL09XU (BL8W) beamline at SPring-8, $\mathcal I^*_{\rm M} \approx 5\times 10^{13}\;{\rm s^{-1} meV^{-1}}$ ($\mathcal I^{**} \approx 1.7\times 10^4 \;{\rm s^{-1} meV^{-1}}$) and $A\approx 3.4\times 10^4$, indicating that some advantage may exist. 

States $\rm N^{**}$ with lower energies exist in $^{57}_{26}$Fe (136~keV), $^{73}_{32}$Ge (67~keV) and $^{181}_{73}$Ta (136~keV), although more work is needed to understand the strength of the transitions to N and $\rm N^*_M$.

\section{Reach}
\label{sec:reach}
 In this section, the reach of the method discussed in probing a variety of scalar and tensor forces is evaluated. The following interactions are considered: 

\begin{equation}
\mathcal{L} \supset y_q \phi \bar{q} q + \frac{\phi}{f_{\gamma}} F_{\mu \nu}^2 + \frac{\phi}{f_g} G_{\mu \nu}^2 + \frac{\tilde{h}_{\mu \nu}}{f_T}{F^{\mu}}_{\sigma}F^{\nu \sigma} + g \phi h^2
\label{Eqn:Moduli}
\end{equation}
where $\phi$ and $\tilde{h}_{\mu \nu}$ are new scalar and tensor interactions, respectively, that couple to quarks ($q$), electromagnetism ($F_{\mu \nu}$), gluons ($G_{\mu \nu}$) and the Higgs ($h$). In the setup shown in Figure~\ref{fig:sketch} where the attractor and the absorber are parallel thin plates separated by a distance $d$, the attractor produces a potential whose value at the absorber is: 

\begin{equation}
    \phi, \tilde{h}_{\mu\nu} \approx \frac{g_N n d^2}{2 e}
    \label{Eqn:field}
\end{equation}
where $g_N$ is the coupling per-nucleon and $n$ the number density of nucleons in the material. This expression assumes that the absorber is placed at a distance $d = \lambda$ where $\lambda$ is the range of the new force, causing the potential to drop by $e^{-d/\lambda} = e$. The scalar $\phi$ shifts nuclear energy levels in the absorber, changing the frequency of the M\"ossbauer line. The tensor $\tilde{h}_{\mu \nu}$ does not significantly change nuclear energy levels. Instead, it changes the frequency of the emitted $\gamma$ as it propagates out of the source towards the absorber. In the following, we estimate these effects (sub-sections \ref{subsec:scalar} and \ref{subsec:tensor})  and compute the reach (sub-section \ref{subsec:sensitivity}) of the experiment.

\subsection{Scalar}
\label{subsec:scalar}

For each of the scalar couplings in Eq~\eqref{Eqn:Moduli}, the effective coupling $g_N$ between a nucleon and $\phi$ as well as the energy shift $\Delta E$ induced in the nucleus due to $\phi$ need to be estimated. Because of the non-perturbative nature of nuclear physics, these effects cannot be calculated from first principles, and, instead, are derived from phenomenological models. These should be regarded as providing the rough order of magnitude of the effect. Generally, the estimates of $g_N$ are more reliable than those of $\Delta E$. Since this is an experiment where a new field is produced and detect, the dependence of the fundamental lagrangian parameters ($y_d,g, f_{\gamma}, f_{g}, f_T$) on $g_N$ and $\Delta E$ is $ y_d,g,  f_{\gamma}^{-1}, f_g^{-1}, f_{T}^{-1} \propto \sqrt{g_N \Delta E}$.

\subsubsection{Yukawa Modulus}

As an example of the Yukawa couplings in Eq.~\eqref{Eqn:Moduli}, we focus on $\phi$ that couples to the down quark. Given the term $y_d \phi \bar{d} d$, the effective nucleon coupling $g_N \phi \bar{N} N$ can be estimated from lattice methods, yielding $g_N \approx 9.5 y_d$ \cite{Hoferichter:2017olk}.

To estimate the energy shift in the nuclear levels due to $\phi$, we calculate the change in the 1-pion exchange potential and equate it to the energy shift $\Delta E$. Using the parametrization in~\cite{Hitoshi}, 

\begin{equation}
   \Delta E =  \Delta V_{\pi} \approx \frac{15}{4} \frac{m_{\pi}^2}{m_N^2}\frac{1}{r} \approx 2.5 y_d \phi ,
\end{equation}
where the relationship between the pion mass and is assumed the down quark mass is used, and is assumed that nuclear level transitions change the relative distance between nucleons (of mass $m_N$) by $r \approx 1$ fm. 

\subsubsection{Alpha Modulus} 

The coupling $\frac{\phi}{f_{\gamma}}F_{\mu \nu}^2$ shifts the fine structure constant $\alpha \rightarrow \alpha - \frac{\alpha^2 \phi}{f_{\gamma}}$. To derive an effective nucleon coupling $g_N$, we compute the contribution of the electromagnetic field to the mass of the nucleus. In a nucleus with $Z$ protons and $A$ nucleons, $\phi$ shifts the self energy of the electromagnetic field by  $-\frac{Z^2 \alpha^2 \phi}{A^{1/3}f_{\gamma}r_p}$, where $r_p \approx 1.2 $ fm is the proton radius. From this, the coupling per-nucleon $g_N = \frac{Z^2 \alpha^2}{A^{4/3}f_{\gamma}r_p}$ is extracted. 

Next, the energy shift $\Delta E$ of the nuclear level caused by $\phi$ is estimated.  M\"ossbauer nuclei such as $^{57}$Fe and $^{181}$Ta have nearly degenerate low lying energy levels implying that the nuclei are highly deformed. Further, their intrinsic quadrupole moments are comparable to $\left(A^{1/3} r_p\right)^2$. The M\"ossbauer transitions are between states that have different spins. Thus, their quadrupole moments are $\mathcal{O}\left(1\right)$ different and it is reasonable to assume that the effective location of the transitioning nucleon changes by the size of the nucleus itself. $\Delta E$ is taken to be equal to the change in the Coulomb energy, yielding $\Delta E = - \frac{Z \alpha^2}{A^{1/3}f_{\gamma}r_{p}} \phi$. 

\begin{figure}
\centering
\includegraphics[width=0.95 \linewidth]{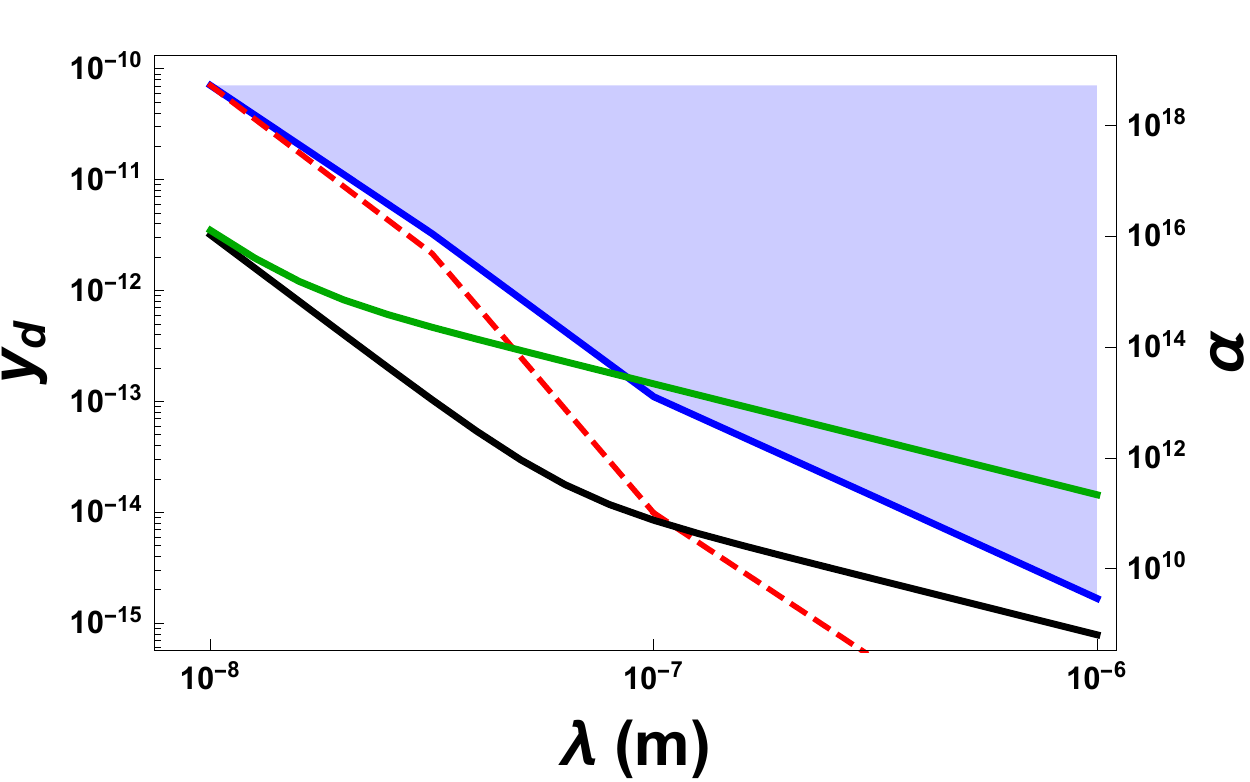}
\caption{Sensitivity to the down quark Yukawa modulus $y_d$, as a function of the range, $\lambda$. The green and the black lines are the projected sensitivities for  $^{57}$Fe ($10^{13}$ total decays) and $^{181}$Ta ($3 \times 10^{14}$ total decays) respectively, assuming natural line width in both cases.      For comparison, the corresponding value of $\alpha$ (strength of force relative to gravity) is shown on the right. The envelope of current limits is indicated by the blue region and the dashed red line, representing the limit obtained in~\cite{IUPUI} where a large background from Casimir interactions was subtracted.}
\label{yukawa}
\end{figure} 

\begin{figure}
\centering
\includegraphics[width=0.95 \linewidth]{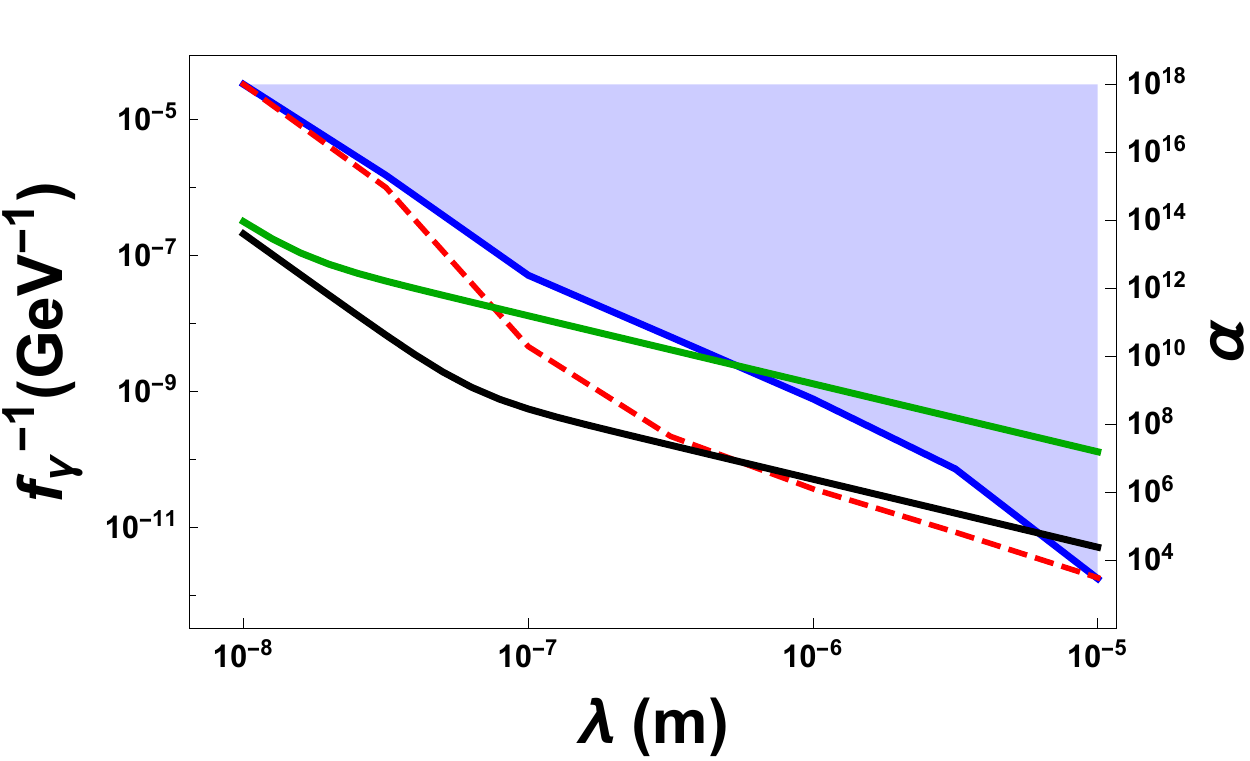}
\caption{Sensitivity to the alpha modulus $f_{\gamma}$, as a function of the range, $\lambda$. The remainder of the description is the same as in Figure~\ref{yukawa}.}

\label{alpha}
\end{figure}

\subsubsection{Gluon Modulus}

The coupling $\frac{\phi}{f_{g}}G_{\mu \nu}^2$ shifts the QCD structure constant $\alpha_s \rightarrow \alpha_g - \frac{\alpha_s^2 \phi}{f_{g}}$. We use the rough argument that the mass $m_N$ of a nucleon should be $m_N \approx \frac{\alpha_s}{r_p}$. This yields $\alpha_s \approx 5$ and the effective nucleon coupling $g_N \approx \frac{\alpha_s^2}{f_{g} r_p}$. To estimate the energy shift in the nuclear levels, the 1-pion exchange potential $V_{\pi} = \frac{g^2}{4} \frac{m_{\pi}^2}{m_N^2} \frac{1}{r_p}$ is used. The change in $\alpha_s$ shifts $g^2$ by $\frac{4 \pi \alpha_s^2 \phi}{f_g}$ yielding an energy shift $\Delta E \approx \frac{0.08 \text{ GeV}}{f_{g}} \phi$.

\subsubsection{Higgs Portal}

The Higgs portal coupling  $g \phi h^2$ emerges naturally in theories such as the relaxion that can solve the hierarchy problem \cite{Graham:2015cka, Fonseca:2018xzp, Budnik:2020nwz}. This coupling shifts nucleon energy levels since it effectively acts as a down quark Yukawa modulus, leading to an energy shift $\Delta E \approx \frac{2.5 g m_d}{m_h^2}\phi$. Its effective coupling to nucleons, $g_N$,  can be estimated via lattice methods and is $g_N \approx 10^{-5} \frac{g}{\text{GeV}}$ \cite{Graham:2015ifn}.

\subsection{Tensor}
\label{subsec:tensor}
The tensor $\tilde{h}_{\mu \nu}$ in \eqref{Eqn:Moduli} couples to the electromagnetic contribution to the mass of the nucleus.  From this mass $\frac{Z^2 \alpha}{A^{1/3}r_p}$, we calculate the coupling per-nucleon $g_N = \frac{Z^2 \alpha}{A^{4/3}f_{T}r_p}$. When a photon of energy $\omega$ traverses a potential difference $\Delta \tilde{h}$,  its energy changes by $\Delta E = \omega \frac{\Delta \tilde{h}}{f_T}$, where, for the present purpose $\Delta \tilde{h} = \tilde{h}$.

\begin{figure}
\centering
\includegraphics[width=0.95 \linewidth]{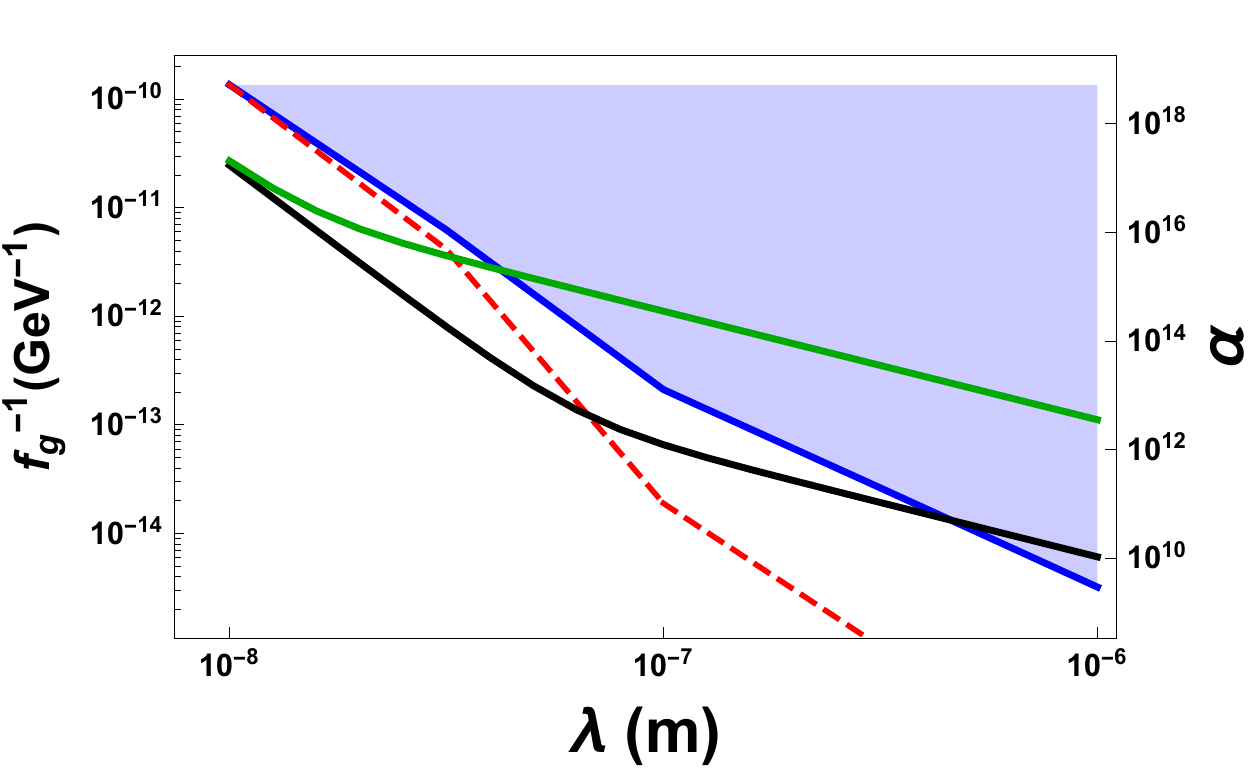}
    \caption{Sensitivity to the gluon modulus $f_g$, as a function of the range, $\lambda$.  The remainder of the description is the same as in Figure~\ref{yukawa}.}
\label{gluon}
\end{figure} 

\begin{figure}
\centering
\includegraphics[width=0.95 \linewidth]{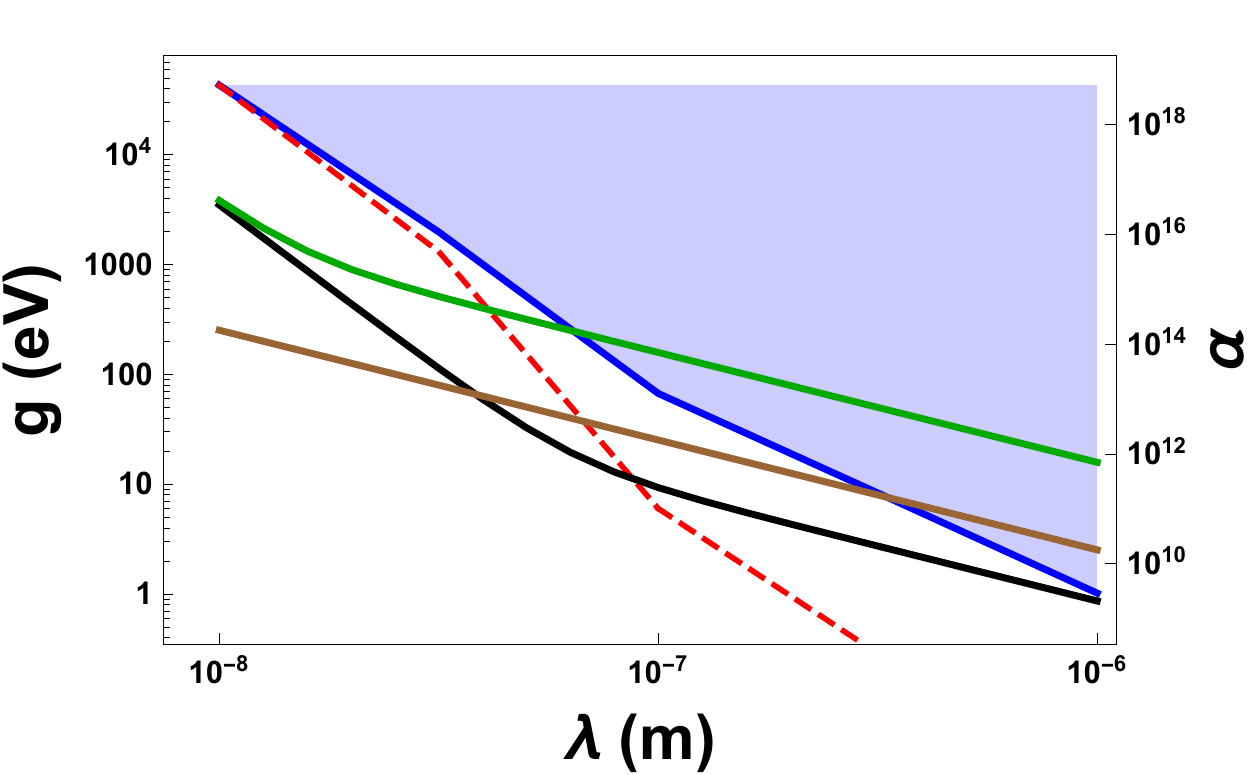}
\caption{Sensitivity to the Higgs portal $g$, as a function of the range, $\lambda$.   The remainder of the description is the same as in Figure~\ref{yukawa}.}
\label{higgs}
\end{figure}

\subsection{Sensitivity}
\label{subsec:sensitivity}

Using Eq.~\eqref{Eqn:field} for the scalar and tensor fields produced, the energy shifts for the various terms in Eq.~\eqref{Eqn:Moduli} are computed using the approximations discussed above. The reach of the experiment for each coupling is calculated as $\Delta E = \frac{\Gamma}{\sqrt{N_{\gamma}}} + 10^{-14}  \left(\frac{10 \text{ nm}}{d}\right)^4 \text{ eV}$. The second term in this expression arises from the systematic limit which is dominated by the chemical shift caused by Casimir forces, as discussed in Sec.~\ref{sec:concept}.  The sensitivity of the technique is estimated disregarding the statistical fluctuations due to possible non-resonant absorption, as the resonant photon statistics can be  obtained by detecting the re-emission of the M\"ossbauer photons by the absorber or through the use of pulsed synchrotron excitation schemes, at least for the short lifetime cases.

$^{57}$Fe and $^{181}$Ta are used as examples, assuming natural line-width are achieved. For $^{57}$Fe, that is taken as the conservative case, we assume a total of $10^{13}$ decays which can be obtained with a commercial 100~mCi source operating for a month, with $\pm 10^{\circ}$ angular collimation. $^{181}$Ta is treated as a more aggressive case, assuming a 1~Ci source operating for a month with $\mathcal{O}\left(1\right)$ angular coverage, such as may be realistic using magnetic tuning.  This yields a total of $3 \times 10^{14}$ decays. In both cases the parameters $\beta$ and $\eta$ from Table~\ref{tab:Traditional} are used; in addition, it is assumed that the total number of decays is evenly split between times when the attractor, made out of gold, is near and far from the absorber.  The corresponding smallest measurable energy shifts are $\Delta E \approx 10^{-15}$~eV for $^{57}$Fe and $\Delta E \approx 10^{-17}$~eV for $^{181}$Ta.   The sensitivity is limited by the chemical shift due to Casimir interactions at distances around 10~nm and becomes statistics limited at larger distances. 

\begin{figure}
\centering
\includegraphics[width=0.95 \linewidth]{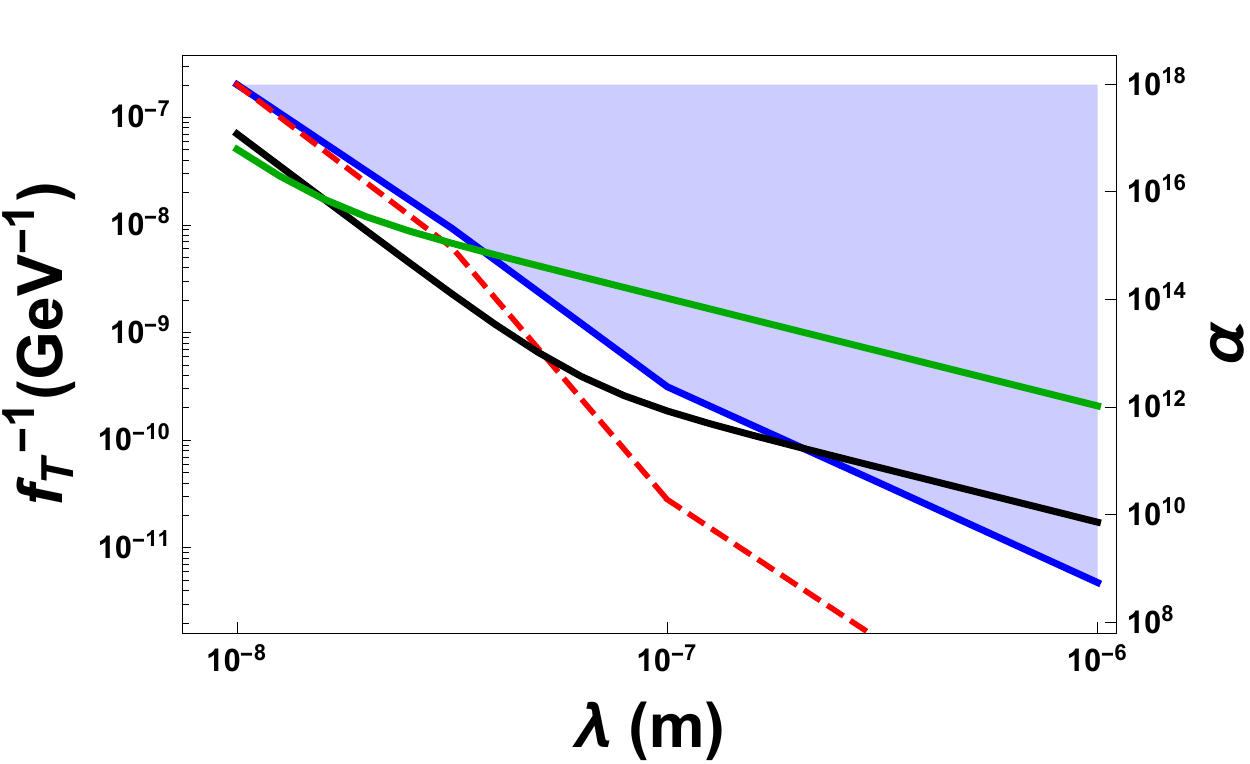}
\caption{Sensitivity to a new tensor force between matter and light, $f_T$, as a function of the range, $\lambda$.  The remainder of the description is the same as in Figure~\ref{yukawa}.}
\label{tensor}
\end{figure}

These results are plotted in Figures~\ref{yukawa},~\ref{alpha},~\ref{gluon},~\ref{higgs} and~\ref{tensor} for the down quark Yukawa, alpha and gluon moduli, Higgs portal and the tensor coupling, respectively. In each figure, the solid green and black lines represents the reach for $^{57}$Fe and $^{181}$Ta, respectively.  The solid blue and red dashed lines represent the current experimental limits on the relevant couplings. The red dashed line is the limit from obtained in~\cite{IUPUI} where a large electromagnetic background from Casimir interactions was subtracted using a common Casimir shield. The solid blue line is the limit without this background subtraction.   For each plot, the strength relative to gravity, $\alpha$ is also shown to the right.  The conversion to $\alpha$ of the parameter $g_N$ used in the computations of $y_d$ (Figure~\ref{yukawa}), $f_g$ (Figure~\ref{gluon}) and $g$ (Figure~\ref{higgs}) is straight forward, since those depend solely on QCD.  The analogous conversion for $f_\gamma$ (Figure~\ref{alpha}) and $f_T$ (Figure~\ref{tensor}) also depends on the atomic number of the nucleus.  For a given nucleus this is a well defined relationship and thus our projected sensitivities can be accurately calculated. But, to convert the limits in~\cite{IUPUI} in terms of our parameters, detailed knowledge of the nuclei used in that setup is necessary. We took the parameters of silicon, as a representative example. 

As is evident from these figures, the M\"ossbauer approach can substantially improve the sensitivity to short distance forces in the range $10^{-8}$m - $10^{-7}$m, beyond current laboratory detection schemes. It can also significantly extend constraints on forces in the range $10^{-7}$m - $10^{-6}$m with a robust natural background suppression instead of relying on the subtraction of large backgrounds. 

With the continuous evolution of synchrotron radiation sources and the possibility of implementing novel excitation schemes (see section \ref{sec:absorption}), it may be possible to access a newer class of M\"ossbauer nuclei (see Table \ref{tab:properties}). It is interesting to speculate on the possible reach of such systems, for example considering the possibility that the 12~keV level of $^{45}$Sc is directly excited by synchrotron radiation. Using the estimates in Sec.~\ref{sec:coherent}, $\approx 10^{5}$ M\"ossbauer photons are expected, yielding an energy sensitivity  $\approx 10^{-17}$~eV, comparable to that of the projected sensitivity of $^{181}$Ta in the above figures. The three level synchrotron excitation scheme could potentially be used to access the 30~keV level of  $^{189}$Os, resulting in as many as $\approx 10^2$  M\"ossbauer photons, corresponding to sensitivity to energy shifts as small as  $10^{-21}$~eV. If successfully developed, this excitation scheme could place competitive limits in the micron range, where the ability of Casimir and patch potentials to shift the nuclear lines are sufficiently suppressed. 

These results also show that there are interesting differences in the potency of this M\"ossbauer approach in relation to the direct measurement of forces between matter as a way to search for new interactions. The M\"ossbauer technique is sensitive both to how the new interaction couples to a nucleon (and thus searched for in measurement of forces) and to the way the new interaction changes nuclear levels. We have already commented on how the latter fact prevents this approach from being sensitive to new vector interactions. This fact is also of importance to scalar interactions since different scalars change the nuclear level splittings differently. For example, the Yukawa and alpha moduli are more efficient in changing the nuclear energy differences as opposed to the gluon modulus whose effects are suppressed by ratios of the pion and nucleon masses. Thus, in the event that a new interaction is discovered, this method provides a unique way to probe the microphysics of the new interaction. 

In addition to these laboratory probes, there are also astrophysical limits on the light particles considered, arising from the possibility that such particles may be emitted by astrophysical systems causing them to cool more rapidly than observed~\cite{Raffelt:1998hw}. Roughly, these limits are at $y_d \approx 10^{-13}$, $g \approx 10$~eV, $f_g \approx 10^{9}$~GeV and $f_{\gamma}, f_T \approx 10^{10}$~GeV. However, these limits are not robust to perturbations of the underlying model - in particular, if there are additional highly suppressed but long-ranged interactions between the standard model and these particles, the astrophysical limits can completely disappear~\cite{DeRocco:2020xdt}. Similarly, density dependent effects can also eliminate these limits~\cite{Masso:2005ym}. Given the very different nature of stellar environments and a terrestrial experiment, there are no model independent astrophysical limits in this part of parameter space.

\begin{comment}
\begin{figure}
\centering
\includegraphics[width=0.95 \linewidth]{Scyukawamodulus.pdf}
\caption{The thick brown line is the sensitivity to the down quark Yukawa modulus $y_d$ from a possible $^{45}$Sc experiment.  The remainder of the description is the same as in Figure~\ref{yukawa}.}
\label{Scyukawa}
\end{figure} 
\end{comment}

\section{Conclusions}
\label{sec:conclusions} 

While, over the years, the M\"ossbauer effect has had a tremendous impact in the fields of material science and chemistry, except for the notable case of the classic experiment verifying the frequency shift of photons in the Earth's gravitational field ~\cite{PoundRebka, KatilaRiski}, it has not had significant applications to the physics of fundamental interactions. The comparison with sensing platforms based on optical and atom interferometers may be of interest.   While the intrinsic  energy resolution of a single M\"ossbauer transition is substantially higher than that of atoms, the much smaller cross sections of nuclear transition results in modest rates.   In addition, resonant techniques are substantially more complex and less explored in the case of nuclear physics, because of the relative lack of suitable instrumentation.  Thus, the M\"ossbauer effect can be competitive only in sensing applications where atom-based sensors encounter difficulties. This is the case for material science and chemistry where the M\"ossbauer effect is used to study nano-scale properties of lattices. The search for short distance forces falls into the same category - substantial electromagnetic effects at short distances inhibits the use of atom-based sensors. The secluded nature of the nucleus inside an electron cloud, along with the suppressed interactions between nuclear moments and electromagnetism, makes the  M\"ossbauer effect ideal for probing short distance forces. 

The requirements for this application is different from those needed for material science and chemistry. In the latter cases,  substantial line shifts are produced and the use of different matrices for source and absorber is required.   For applications to fundamental interactions the interest is in obtaining the narrowest possible lines and the smallest possible intrinsic shift between source and absorber. 

Excitingly, we have shown that a first round of competitive measurements is possible using the traditional M\"ossbauer technique as well as nuclear resonance with synchrotron radiation excitation.  Further technical improvements in both areas may extend the sensitivity well beyond the current state of the art.   At the shortest distances, the M\"ossbauer approach presented here appears to be limited by chemical shifts due to Casimir interactions. However, since this approach is tailored to the discovery of scalar interactions by measuring a line shift rather than the displacement of an object, it might be possible to mitigate this background. For example, one may consider placing the absorber between two attractors, measuring the distance between the absorber and the attractors and feeding back to null out the Casimir force on the absorber. Unlike for the case of conventional force sensing experiments, a new scalar interaction from the attractors will simply add, further shifting the line.  Since the background is a second order effect, even a partial cancellation results in substantial improvements.

It may also be possible to mitigate the background by adopting a differential measurement with two different nuclides, e.g. $^{57}$Fe and $^{181}$Ta. The absorber could be made of an alloy/compound containing both, so that both are subject to the same local molecular structure affected by the electric coupling to the attractor.  The line shift produced by the new force is generally expected to be different from that originating from the change in molecular fields for the two nuclides, so that, effectively, one nuclide can be used as a ``co-electrometer'' to measure the effect of electric fields, separately from the new interaction.  The ability to potentially implement such a differential measurement strategy is a key difference between this approach and conventional force measurements. In the latter case, once electromagnetic backgrounds dominate, the ability to look for a new interaction is effectively blocked. With the  M\"ossbauer setup, it is possible to get additional information, potentially providing a path towards better sensitivity. 

\section{Acknowledgements}

One of us (GG) is grateful to E.E.~Alp (ANL) for an introduction to synchrotron radiation excitation of M\"ossbauer states and to U.~Bergmann, A.~Halavanau, C.~Pellegrini (SLAC) for discussions on X-ray lasers.  We  also gratefully acknowledge some preliminary discussions with J.~Schiffer (ANL) and the help  of P.~Vogel (Caltech) in understanding several nuclear physics phenomena.   Finally, we thank E.E.~Alp and P.~Vogel for their careful reading of the manuscript.

\end{document}